\documentstyle[12pt,epsf]{article}
\newcommand{\spav}[1]{\parbox{1mm}{\vspace*{#1}}}

\begin{document}
\begin{titlepage}
\begin{center}
{\Large\bf 
Microlensing searches of dark matter}
\spav{1cm}\\
{\large Esteban Roulet}\spav{1.5cm}\\
{\em Depto. de F\'\i sica, Universidad Nacional de La Plata}\\ {\em 
CC67, 1900, Argentina.}\spav{1.5cm}\\
{\sc Abstract}
\end{center}

The evolution of the observational results of microlensing towards the
 LMC and some of the suggested interpretations to account for them are
 discussed. It is emphasized that the results at present are indicative of
 a lensing population of white dwarfs, possibly in the spheroid (not
 dark halo) of the Galaxy, together with the more standard backgrounds
 of stellar populations in the Magellanic Clouds and in the
 Galaxy. This is also hinted by dynamical estimates of the spheroid
 mass and by recent direct searches of old white dwarfs. 
 \vfill

\end{titlepage}

\newpage
\setcounter{footnote}{0}
\setcounter{page}{1}

\section{Introduction:}

To understand the nature of the dark matter is one of the
pressing unsolved mysteries in cosmology and particle physics
nowadays.  The dark matter problem is the fact that the matter in
luminous forms (stars and gas) is inferred to account only for
$\Omega_{lum}\simeq 0.007$, i.e. for less than one percent of the
critical density. On the other hand, dynamical estimates based on the
gravitational influence of the mass on test bodies (gas orbiting
galaxies, galaxies moving inside clusters) imply that there  is at
least ten times more `dark' mass than luminous one on galactic scales,
and even more ($\Omega\sim 0.3$) on cluster scales. The issue on even
larger scales is still unsettled, existing at present indications in
favor of some dark energy (cosmological constant) accounting for
$\Omega_\Lambda\simeq 0.7$, i.e. filling the gap up to the inflation
preferred value  $\Omega=1$ corresponding to a flat universe. 

The
possible constituents of the dark matter naturally split into baryonic
and non-baryonic candidates. The last ones would be some kind of
weakly interacting particle permeating the universe, such as
supersymmetric neutralinos, axions or massive neutrinos, while the
first ones would be made of just ordinary protons and neutrons hidden
in some non-luminous forms. The simplest examples to achieve this
would be to have them in MACHOs (the acronym for 
massive astrophysical compact halo objects),
 such as  stellar remnants (white dwarfs, neutron stars, black holes),
brown dwarfs (stars with mass $m<0.1M_\odot$, which never become hot
enough to start sustained nuclear fusion reactions) or even
planets. Also cold gas clouds have been suggested as dark baryonic
constituents of the galactic halos.

The main constraints on the amount of baryons in the universe  come
from the theory of primordial nucleosynthesis, which requires
$\Omega_b\simeq 0.01$--$0.05/(H_0/70$ km/s/Mpc)$^2$, with the lower
(higher) values corresponding  to the high (low) primordial deuterium
abundance determinations from QSO absorption lines \cite{ol99}. 
Hence, we see that
for a Hubble constant $H_0\simeq 50$--70~km/s/Mpc nucleosynthesis
indicates that dark baryons should exist and that their total amount
is  probably insufficient  to account for dark
galactic halos ($\Omega\simeq 0.1$). 
Clearly to account for the observations on
cluster scales non-baryonic dark matter is also required, and it would
then be natural to expect to have some amount of it also at galactic
scales. 

Although dark compact objects are, by definition, very hard to be seen
directly, they may reveal themselves through the gravitational lensing
effect they produce on background stars. Indeed in 1986 Paczynski
\cite{pa86} 
showed that by monitoring a few million stars in the Large Magellanic
Cloud (LMC) during a few years would allow to determine whether  the dark
halo of the Galaxy consists of MACHOs with masses in the range
$10^{-7}$--$10^2\ M_\odot$, covering essentially  all the range of
suggested candidates. Soon after, with the advent of CCDs this program
became feasible and several groups started the searches at the
beginning of the nineties.

\section{Microlensing expectations:}

The gravitational deflection of light by massive bodies is one of the
predictions of general relativity which is now tested to better than
the percent level. This phenomenon leads to several beautiful effects
such as the multiple imaging of QSOs by intervening galaxies, giant
arcs around clusters or weak lensing distortions of faint background
galaxies, which have the potential of giving crucial information for
cosmology (measurement of $H_0$ from the time delays between different
QSO images, estimates of cluster masses, etc.). When the deflector is
a compact star like  object, a MACHO, two images of the background
sources are produced, one on each side of the deflector. In the
particular case of perfect lens-source alignment, the image is a ring
with angular size (Einstein's angle) 
$$\theta_E\simeq {\rm mas}\sqrt{{m\over M_\odot}{(1-x)\ 10{\rm
kpc}\over D_{o\ell}}},$$
where taking $D_{os}$ as the distance from the observer to the source,
$D_{o\ell}\equiv xD_{os}$ is the distance to the lens. Hence, for lenses
at a few kpc distance with masses in the range of interest for the
MACHO searches this angle is much smaller than the typical telescope
resolution ($\sim 40$~mas for HST!) and hence cannot be
resolved. However, the important effect is that the brightness of the
source  is magnified due to a lensing effect (for reviews see
refs.~\cite{ro97,pa96}) by a total amount
$$A={u^2+2\over u\sqrt{u^2+4}}\label{ampli}$$
where  $u^2=\xi^2/R_E^2$, with $R_E=\theta_ED_{o\ell}$ being the radius of
the Einstein ring in the lens plane and $\xi$ is the distance between
the lens and the line of sight to the source. Due to the lens relative
motion orthogonally  to the l.o.s. with velocity $v^\perp$, one has
$\xi^2=b^2+v^{\perp 2}(t-t_0)^2$, with $b$ the impact parameter
of the lens trajectory and $t_0$ the time of closest approach. Hence,
the amplification will vary with time in a very specific form and it
is sizeable ($A>1.34$) as long as the lens distance to the line of
sight is not larger than $R_E$.
 Hence, one can estimate the optical depth,
i.e. the probability  that a given star is magnified significantly for
a given lens distribution with density $n_\ell$, as the number of
lenses within a distance $R_E$ from the l.o.s.. If 
the lenses are assumed to have a common mass $m$, one has
$n_\ell=\rho_\ell/m$, and hence 
$$\tau={4\pi \over m}\int_0^{D_{os}}{\rm d}D_{o\ell}\ R_E^2 \rho_\ell.$$
 In the case of a halo consisting fully of
MACHOs, the lens density will be just the halo density, which from the
observed rotation curve of the Galaxy should be 
$$\rho_H\simeq \rho_0^H{r_0^2+a^2\over r^2+a^2},$$ 
with $r_0=8.5$~kpc the distance to the galactic center, the core
radius $a$ is of a few kpc and the local halo density is
$\rho_0^H\simeq 10^{-2}M_\odot$/pc$^3$. This would lead to a predicted
optical depth for LMC stars of $\tau^H\simeq 4\times 10^{-7}$.

The timescale of the events, defined as $T\equiv R_E/v^\perp$  (and
determined from a fit to the light curve with the theoretical
expression for $A(u)$)
has an average value $\langle T\rangle \simeq 65 {\rm
d}\sqrt{m/M_\odot}$ given the typical velocity dispersion expected for
halo objects ($\sigma\simeq 155$~km/s). One can show that for 100\%
efficient searches the number of expected events is
$N=(2/\pi)(\tau/\langle T\rangle)\times {\rm Exposure}\simeq 20 \ {\rm
events} 
\sqrt{M_\odot/m}({\rm Exposure}/10^7{\rm\  stars\ yr})$. Actually, the 
efficiencies are $\leq 30\%$ and are  time dependent,
 so that a detailed prediction requires to
convolute the differential rates with the efficiencies
(and also with the mass distribution of the lenses).

\section{First microlensing results and their interpretation}

By 1992 the observational searches towards the LMC by the french EROS
and australo-american MACHO collaborations had started and already
after one year the first candidate events had showed up. It soon
became apparent however that the number of events observed were a
factor $\sim 5$ below the expectations from a halo made of objects
lighter than a few solar masses. The estimate of the optical depth to
the LMC from the first 3 MACHO events \cite{al96} was indeed
$\tau= 8.8^{+7}_{-5}
\times 10^{-8}$ and the one from the first two EROS events \cite{re97}
was similar,
$\tau=8.2^{+11}_{-5\ } \times 10^{-8}$. 
The typical event durations of a few weeks were also
indicative of lens masses in the ballpark of 0.1~$M_\odot$, i.e. in
the limit between brown dwarfs and very light main sequence
stars. Furthermore, the non observation of short duration events ($T$
less than a few days) by the EROS CCD program and by MACHO set
stringent bounds on light lenses ($10^{-7}<m/M_\odot<10^{-3}$) which
could contribute no more than $\sim 25$\% to the overall halo mass
\cite{al98}. 

Although the number of events were few, they were significantly larger
than the backgrounds expected from the known stellar populations. For
instance, the faint stars in the galactic disc would contribute little
to the optical depth towards the LMC, because although the local disc
density is an order of magnitude larger  than the halo one
($\rho^D_0\simeq 10^{-1}M_\odot/$pc$^3$) its scale height is very
small ($h\sim 100$--300~pc). In general, adding an exponential disc
like population with local column density $\Sigma$ leads to an optical
depth to the LMC of $\tau^D\sim
10^{-8}(\Sigma/30M_\odot/$pc$^2)(h/300$~pc) \cite{go94}. 
The column density in the
thin disc is $\sim 50M_\odot/$pc$^2$, of which 20\% is in
gas. Furthermore taking into account the measured mass function of
disc stars it can be shown that a large fraction of $\tau^D$ would be
due to stars too bright to go undetected \cite{al00}, 
and hence the thin disc
predicted depth is actually below $10^{-8}$. 

The Galaxy is also known to have a thicker disc of stars, with
$h\simeq 1$~kpc, probably originating from a past merger of a
satellite which somehow heated the disc stars. The column density
associated to the observed thick disc stars is typically taken as
$\Sigma^{TD}\sim 4M_\odot/$pc$^2$, and hence the expected optical
depth is tiny ($\sim 0.5\times 10^{-8}$). However, in
ref. \cite{go94}
it was noticed that the maximum allowed column density consistent
with dynamical observations, $\Sigma^{TD}_{max}\simeq
50M_\odot/$pc$^2$, would lead to $\tau\simeq 5\times 10^{-8}(h/$kpc),
and hence for the scale height adopted there (1.4 kpc) it would already
be close to the observed value.

Another known stellar galactic population is the spheroid (or `stellar
halo'), observed through the old metal poor stars at high latitudes
and also as high velocity stars nearby (the observed
velocity dispersion is $\sigma\simeq 120$~km/s). 
It was formed in the first
Gyrs of the Galaxy lifetime, has a probably slightly flattened
spherical shape ($c/a\simeq 0.7$--1)with density profile
$r^{-\alpha}$, with $\alpha\simeq 2.5$--3.5 \cite{ro00}. 
The predicted optical depth for this population, adopting $c/a=1$ and
$\alpha=3.5$, is $\tau^S\simeq 0.3\times 10^{-8}
(\rho_0^S/10^{-4}M_\odot/$pc$^3$). This value would increase by 50\%
for $\alpha=2.5$ and decrease by $\sim 25$\% for $c/a=0.7$. The
spheroid density inferred from star counts, $\rho_0^S\simeq 5\times
10^{-5}M_\odot/$pc$^3$ leads then to a tiny optical depth. However,
there have been since many years suggestions that the spheroid density
should be larger by an order of magnitude, based on global fits to the
galactic rotation curve and other dynamical observations \cite{os82}. 
In these
heavy spheroid models the optical depth results $\sim 4\times
10^{-8}$, and is hence relevant for observations towards the LMC \cite{gi94}.
 
Another background for LMC searches is that of old neutron stars
\cite{mo97}. These are born in the disc, but they are believed to 
acquire large
velocities in the supernova explosions producing them and hence move
to large distances where they become more efficient lenses. For the
typical value $N_{ns}\simeq 2\times 10^9$ for the number of past
galactic core collapse supernovae \cite{ti96}, 
the resulting optical depth is $\tau^{NS}\simeq
0.4\times 10^{-8}$, comparable to the one from known stars in the
spheroid or thick disc.

Regarding the LMC populations, in the same way as the Milky Way has a
dark halo, the rotation curve of the LMC suggests that this galaxy
also has a
dark halo around it. The optical depth associated to it amounts
to $\sim 8\times 10^{-8}$ \cite{go93,wu94}. Hence, if all the dark matter
were in MACHOs the total optical depth would be 20\% higher than the
prediction from the Milky Way halo alone, and this clearly worsens
the  discrepancies with the observations. It has to be stressed that
no stellar population has been observed with the expected LMC halo
distribution. The predictions for the observed distributions of LMC
stars (disc and bar) are more delicate. It was actually suggested that
the optical depth towards the central bar could be $\sim 5\times
10^{-8}$ \cite{sa94}, and hence of the order of the observed
$\tau$. However, outside the bar, where several of the observed events
actually are, $\tau$ falls significantly, and the most recent
estimates for the expected average depth \cite{gy99} 
are $\tau\simeq 2.4\times
10^{-8}$, although with some spread among different models.

Summarizing, the expectations from known stellar populations amount
to $\tau\simeq 3$--4$\times 10^{-8}$, while from the Milky Way and LMC
halos one would expect $\tau\simeq 5\times 10^{-7}$. The observed
value $\sim 8\times 10^{-8}$ was clearly larger than the first one,
but only a small fraction of the second. Since the main purpose of the
microlensing searches was to establish whether the halo consisted of
MACHOs, the results are most commonly presented  in terms of the
fraction $f$ of the Milky Way halo in the form of compact objects, which
for the observed $\tau$ would correspond to $f\simeq 20$\%. The
remaining fraction could be just in cold gas clouds or in non-baryonic
dark matter.  An alternative explanation was to have instead a heavy
spheroid or thick disc made mainly of MACHOs or a very large LMC
self-lensing contribution, and instead a completely
non-baryonic halo.

\section{The second period (1996-2000)}
In 1996, the results of the analysis of the first two years of MACHO
data were announced \cite{al97}. A total of 8 candidate events were
observed and the average event duration turned out to be  longer. 
The conclusion
was that now $\tau=2.9^{+1.4}_{-0.9} \times 10^{-7}$ and also that 
the typical lens masses increased to $m\simeq 0.5M_\odot$.
The picture which emerged then was that 50\% of the halo should be in
objects with the characteristic mass of a white dwarf. This scenario was
quite unexpected, and was shown to be also potentially in trouble with
the chemical enrichment of the galaxy due to all the metal pollution
which would result from the white dwarf progenitors \cite{gi97,fi98}. 
 There were also
attempts to explain the  observed rates as due to some tidal debris of the
LMC or even due to a previously undetected satellite galaxy along the
l.o.s. to the LMC \cite{zh98}, but it is not clear whether these ideas
are actually supported by observations \cite{al97b,be98,za99,gr00}.

In the period after 1996 significant improvements on the observational
programs took place: the EROS and OGLE collaborations started to use
better cameras in new telescopes and networks of telescopes were
organized to follow ongoing alerted microlensing events (GMAN, MOA,
PLANET, MPS). This allowed for instance to measure in great detail an
event in the Small Magellanic Cloud caused by a binary lens
\cite{af99}. 
When the
lens is a binary, the signatures are quite different from the single
lens case.  For large impact parameters there are now three
images of the source, one on the exterior side of each of the lenses
and the third one in between the lenses. An extra pair of images
can appear however for small impact parameters when the source crosses
the so called caustic, and then two images disappear again when the
source leaves the region delimited by the caustic. At these caustic
crossings the magnification formally diverges. This means that one has
an extremely powerful magnifying glass to look at the source. Actually,
effects associated to the finite size of the source become 
observable and in particular they limit the maximum amplification to a
finite value. If one knows the radius of the source, one can also 
infer the
relative lens-source proper motion $\mu$, and since this quantity is
expected to differ significantly for lenses in the Magellanic Clouds
($\mu\sim 1$~km/s/kpc) or in the galactic populations  
($\mu\ge 10$~km/s/kpc), it was possible to establish that the binary
lens belonged to the SMC and not to the halo. Also one of the MACHO
LMC events was a binary and its proper motion again suggested that it
belonged to the LMC. This shows that indeed the contribution of the
lenses in the Magellanic Clouds is significant, 
and actually in the SMC it is
expected to be relatively larger than in the LMC due to the elongated
shape of the SMC along the l.o.s.. 

\section{Recent developments}

The most recent microlensing results appeared a few months ago and
again changed the overall picture. The analysis of 5.7 yrs of MACHO
data \cite{al00} in 30 fields (out of a total of 82)  found 13--17 events
(depending on the criteria used), leading to a significantly reduced
depth $\tau= 1.1^{+0.4}_{-0.3} \times 10^{-7}$ (for the 13 event sample).
Another relevant observation was that 
 no particular increase in $\tau$ towards the center of the Cloud
was observed, contrary to the  expectations from 
a rate dominated by LMC stars.
Also the EROS group presented the reanalysis of their old data together
with two new years of EROS II data \cite{la00}, 
finding a total of only 4 events,
a result inconsistent with the '96 MACHO result. This shifted the
situation somehow back to the initial stage in which the observed
rates are a factor $\sim 5$ below the halo predictions but $\sim 3$--4
above the expectations from known populations. The inferred masses
$m\simeq 0.5 M_\odot$ 
continue to suggest however that the objects could be old white
dwarfs.

Another important related result has been the recent activity related
to the direct search of old white dwarfs. Hansen \cite{ha98} realized
that the spectra of old white dwarfs having molecular hydrogen
atmospheres (i.e. probably half of the total) would look much bluer
and brighter  than previously believed due to the strong absorption
at wavelength larger than 1~$\mu$m by their atmospheres.  This prediction
was actually confirmed by the direct measurement  of the spectrum of a
cool ($T\simeq 3800\ ^\circ$K) nearby white dwarf \cite{ho00}.
 With this new scenario the search for old white dwarfs becomes
feasible, and indeed analyses of two Hubble Deep
Fields taken two years apart were done searching for objects with large
proper motions  \cite{ib99}. The candidate  objects they found, 
with colours consistent
with being old `halo' white dwarfs, led them to infer that their
density could be comparable to the local halo density ($\sim
10^{-2}M_\odot/$pc$^3$). However, recently two new searches in larger
nearby volumes (not so deep but wider) have found results inconsistent
with such large white dwarf densities. Flynn et al. \cite{fl00} found
no candidates while expecting a few tens based on ref.~\cite{ib99}, while
Ibata et al. 
found two nearby white dwarfs \cite{ib00}, inferring a
density of $\sim 7\times 10^{-4}M_\odot/$pc$^3$ for these hydrogen
atmosphere white dwarfs. A remarkable thing is that this value is just
in the required range to account for the missing mass of the heavy
spheroid models, whose proper motions would be only $\sim 20$\%
smaller than those assumed for `dark halo' objects and hence
consistent with those found.  To analyse the possibility that these
old white dwarfs are genetically related to the old population II
spheroid stars, and not to an independent halo population with no
observed counterpart, seems then particularly relevant. If this were
the case, the initial mass function of spheroid stars would have to be
peaked  at a few solar masses to account for the large number of white
dwarf progenitors, and the gas released during the ejection of their
envelopes would have ended up in the disc and bulge, but producing
certainly less metal pollution than the halo white dwarf models due to
the much smaller total spheroid mass.
The future searches of white dwarfs should also be able to distinguish
between thick disc and spheroid/halo populations due to their
significantly different proper motions. 

Regarding the future of microlensing observations, the MACHO
experiment has finished taking data, while EROS II and OGLE II are
still running. A significant increase in the number of events is then
expected when all the data available gets analysed. Furthermore, a new
analysis technique (Difference Image Analysis), devised for the study
of lensing in crowded fields such as the Andromeda galaxy, has been
successfully applied to several bulge fields by the MACHO group
\cite{al00b}, doubling the number of observed events with respect to
previous analyses, and also by EROS to their first CCD data
\cite{me98}.
 The use of this technique for all the LMC data should then also
help to get more decent statistics and hence to discriminate among the
population(s) responsible for the microlensing events.
Clearly the possibility that the dark halo is completely made of
non-baryonic dark matter still remains open, and hence the search for
its even more elusive constituents is crucial to finally unravel the
dark matter mystery.


 \section*{Acknowledgments}
This work was supported by  CONICET,  ANPCyT and  Fundaci\'on Antorchas,
Argentina.

\end{document}